\documentclass[a4paper,11pt]{article}
\usepackage{float}
\usepackage[pdftex]{graphicx}
\usepackage{subcaption}
\usepackage[T1]{fontenc}
\usepackage{lmodern}
\usepackage{slashed}
\usepackage[utf8]{inputenc}
\usepackage[english]{babel}
\usepackage{microtype}
\usepackage{cite}
\usepackage{amsmath,amssymb,amsfonts,amsthm}
\usepackage{mathtools,mathrsfs,calligra,aurical}
\usepackage[nottoc,notlot,notlof]{tocbibind}
\usepackage{upgreek}
\usepackage{mathtools}
\numberwithin{equation}{section}
\allowdisplaybreaks
\usepackage[all]{xy}
\usepackage{xcolor}
\usepackage{graphicx}
\graphicspath{{images/}}
\usepackage{geometry}
\usepackage[toc,page]{appendix}
\usepackage{hyperref}
\usepackage[normalem]{ulem}
\hypersetup{colorlinks = true, linkcolor = red, linktocpage = true, citecolor = blue}

\usepackage{bm}
\usepackage{ragged2e}
\usepackage{appendix}
\usepackage{slashed}
\usepackage{bbold}
\usepackage{cancel}
\usepackage{setspace}
\newcommand{\be}{\begin{equation}}
\newcommand{\bea}{\begin{eqnarray}}

\newcommand{\ee}{\end{equation}}
\newcommand{\eea}{\end{eqnarray}}

\DeclareMathAlphabet{\mathpzc}{OT1}{pzc}{m}{it}

\global\long\def\dd{\mathrm{d}}%
\global\long\def\jj{\jmath}%

\global\long\def\bR{\mathbb{R}}%

\global\long\def\Re{\mathrm{Re}}%
\global\long\def\Im{\mathrm{Im}}%
\global\long\def\ri{\mathrm{i}}%

\usepackage{setspace}

\begin{document}

\begin{titlepage}
\begin{flushright}
\par\end{flushright}
\vskip 0.5cm
\begin{center}
\begin{spacing}{2.4}
\textbf{\huge Supersymmetric Moduli Space and Vacua with Vector Fields in $D=4$ Gauged $\mathcal{N}=8$ Supergravity}\\
\end{spacing}

\vskip .5cm

\large {\bf Andr\'{e}s Anabal\'{o}n}$^{~a,~b}$\footnote{anabalo@gmail.com}, \large {\bf Stefano Maurelli}$^{~c,d}$\footnote{stefano.maurelli@polito.it},\vspace{0.4cm}\\ \large {\bf Marcelo Oyarzo}$^{~e}$\footnote{marceloandres.oyarzo@usc.es} and \large {\bf Mario Trigiante}$^{~c,d}$\footnote{mario.trigiante@polito.it}

\vskip .5cm 

$^{(a)}${\textit{Departamento de F\'isica, Universidad de Concepci\'on,\\ Casilla, 160-C, Concepci\'on, Chile.}}\\ \vskip .3cm
$^{(b)}${\textit{Instituto de F\'isica Te\'orica, UNESP-Universidade Estadual Paulista \\
R. Dr. Bento T. Ferraz 271, Bl. II, Sao Paulo 01140-070, SP, Brazil.}}\\ \vskip .3cm
$^{(c)}${\textit{Politecnico di Torino, Dipartimento di Scienza Applicata e Tecnologia,\\ corso Duca degli Abruzzi 24, 10129 Torino, Italy.}}
\\ \vskip .3cm
$^{(d)}${\textit{INFN, Sezione di Torino, Via P. Giuria 1, 10125 Torino, Italy.}}
\\ \vskip .3cm
$^{(e)}${\textit{Departamento de Física de Partículas $\&$ Instituto Galego de Física de Altas Enerxías (IGFAE),
Universidade de Santiago de Compostela, E-15782 Santiago de Compostela, Spain}}
\end{center}

\vskip .5cm 
\begin{abstract}
When fermions are taken to be anti-periodic along a spacelike $S^1$ in $AdS_4$, certain ground states of the supergravity theory are described by AdS-soliton-like spacetimes. We study these geometries within the purely dilatonic STU model obtained from the compactification of M-theory on $S^7$ in the presence of non-trivial gauge fields. The resulting configurations define a rich family of everywhere regular supersymmetric vacua that holographically describe strongly coupled ``confining'' gauge theories in three dimensions. We provide a complete characterization of the moduli space of supersymmetric solutions in terms of the vacuum expectation values of the dimension-one operators of the truncation.
\end{abstract}

\vfill{}
\vspace{1.5cm}
\end{titlepage}

\setcounter{footnote}{0}
\tableofcontents

\section{Introduction}
Anti-de Sitter (AdS) solitons \cite{Horowitz:1998ha} are dual to the ground state of quantum field theories with a compact spatial dimension. Since the associated cycle smoothly closes off in the interior, the absence of singularities enforces anti-periodic conditions on fermionic fields along that direction. In asymptotically flat spacetimes, these boundary conditions typically induce the spontaneous formation of “bubbles of nothing” \cite{Witten:1981gj}. Yet, when suitable background gauge fields are introduced, supersymmetric configurations with anti-periodic boundary conditions can exist \cite{Canfora:2021nca, Nunez:2023nnl, Nunez:2023xgl}. 

When spacetime is asymptotically locally AdS, an extremely simple supersymmetric AdS soliton happens to exist \cite{Anabalon:2021tua}. These solutions describe the confining phase of the dual field theory, and the validity of the supergravity approximation has been investigated extensively in the literature \cite{Fatemiabhari:2024aua,Chatzis:2024top,Chatzis:2024kdu,Giliberti:2024eii,Castellani:2024ial,Macpherson:2024qfi,Barbosa:2024smw,Kumar:2024pcz,Fatemiabhari:2024lct,Jokela:2025cyz,Nunez:2025gxq,Chatzis:2025dnu,Macpherson:2025pqi,Nunez:2025ppd,Chatzis:2025hek,Fatemiabhari:2025usn,Fatemiabhari:2026goj,Nastase:2026lhz,Fatemiabhari:2026rju,Jokela:2026cjm,Chatzis:2026ekd}.  

The existence of domain walls, singular in the infrared, was noticed in the early holographic explorations of the Coulomb branch of different gauge theories \cite{Freedman:1999gk,Cvetic:2000zu,Cvetic:1999xx}. In five dimensions, IR-regular geometries with a single supergravity scalar were constructed in \cite{Anabalon:2024che}, and it was shown that the phase boundary between the different Coulomb branch behaviours corresponds to the point at which the scalar VEV vanishes. Later, the general STU model case was constructed in \cite{Anabalon:2026yxk}. Indeed, the entire family of supersymmetric solutions is parametrized by the scalar expectation values. While the UV boundary conditions determine the asymptotic sources, demanding regularity in the IR imposes additional constraints that fix the scalar VEVs as functions of the sources and the energy scale. Consequently, at each energy scale there exists a moduli space of regular solutions sharing the same UV sources. The submanifold on which the scalar VEVs vanish marks a phase transition of the dual theory, a phenomenon that has also been recently identified in supersymmetric supergravity black holes \cite{Anabalon:2024lgp,Anabalon:2025sok}. In particular, imposing regularity in the IR dynamically selects the allowed vacua, thereby fixing what would otherwise be arbitrary integration constants and promoting them to physically meaningful order parameters of the dual theory. 

Some of us have previously studied the regularization of the Coulomb branch of gauge theories in the various single-scalar truncations of the STU model obtained from the $S^7$ compactification of M-theory to four dimensions \cite{Anabalon:2024qhf,Anabalon:2022aig}. Upon uplifting the supersymmetric solutions to eleven dimensions, we found a precise correspondence between the harmonic functions characterizing the four-dimensional solutions and the Coulomb branch distributions of M2-branes in the eleven-dimensional geometry. In this paper, we solve this problem in full generality within the purely dilatonic sector of the STU model obtained from the $S^7$ reduction of M-theory to four dimensions. For the supersymmetric solutions, we provide a complete characterization of the corresponding moduli space, explicitly construct the Killing spinors, and demonstrate that they are globally well defined and regular throughout the spacetime.

The remainder of the paper is organized as follows. In Section~\ref{sec:STU} we review the purely dilatonic sector of the STU truncation of $D=4$ gauged $\mathcal{N}=8$ supergravity and present the supersymmetry equations relevant for our analysis. In Section~\ref{sec:soliton} we construct the general four-charge soliton solution and characterize its phase space in terms of the asymptotic Wilson lines and the periodicity of the contractible circle. We then focus on the supersymmetric locus, derive the complete moduli space of regular BPS configurations, and express the vacuum expectation values of the dual dimension-one operators in terms of the boundary data. The different regions of the moduli space and the codimension-two loci where the scalar condensates vanish are analyzed in detail. In Section~\ref{sec:susy} we solve the Killing spinor equations explicitly and show that the supersymmetric solutions are globally regular and admit well-defined anti-periodic spinors along the contractible cycle. We conclude in Section~\ref{sec:conclusions} with a discussion of the holographic interpretation of the moduli space, its phase structure, and possible extensions of our analysis. In Appendix~A we provide the general degree-12 polynomial whose roots determine all possible non-supersymmetric solutions of the model.

\section{STU Model} \label{sec:STU}
In this Section, we discuss the construction of an AdS-soliton configuration in the so-called dilatonic sector of the STU model of the gauged $\mathcal{N} = 8$ supergravity. In the dilatonic sector the complex scalars parameterizing the special K\"ahler
manifold are purely imaginary $z_{i}=\ri e^{-\phi_{i}}$, $i=1,2,3$. The effective action principle, which we consider for practical purposes, is given by
\begin{align}
\mathcal{S} & =\int\dd^{4}x\sqrt{-g}\left(\frac{R}{2}-\frac{1}{4}\sum_{i}\partial_{\mu}\phi_{i}\partial^{\mu}\phi_{i}-\frac{1}{4}\sum_{\Lambda}Y_{\Lambda}F_{\mu\nu}^{\Lambda}F^{\Lambda\mu\nu}+\frac{1}{L^{2}}\sum_{i}\cosh\phi_{i}\right)\,.\label{action_dilatonic_STU}
\end{align}
The field equations coming from the action principle (\ref{action_dilatonic_STU})
are given by 

\begin{align*}
\dd(Y_{\Lambda}\star F^{\Lambda}) & =0\,,\hspace{1cm}\Lambda=1,\dots,4\\
\frac{1}{2}\Box\phi_{i}+\frac{1}{L^{2}}\sinh\phi_{i}-\frac{1}{4}\sum_{\Lambda=1}^{4}\frac{\partial Y_{\Lambda}}{\partial\phi_{i}}F_{\mu\nu}^{\Lambda}F^{\Lambda\mu\nu} & =0\,,\hspace{1cm}i=1,2,3\,,\\
R_{\mu\nu}-\frac{1}{2}g_{\mu\nu}R & =T_{\mu\nu}^{(\phi)}+T_{\mu\nu}^{(F)}\,,
\end{align*}
where the energy-momentum tensor of the fields entering in the Einstein's equations and the coupling $Y_\Lambda$ between the vectors and the scalars are given by
\begin{align}
T_{\mu\nu}^{(\phi)} & =\frac{1}{2}\sum_{i=1}^{3}\left[\partial_{\mu}\phi_{i}\partial_{\nu}\phi^{i}-g_{\mu\nu}\left(\frac{1}{2}(\partial\phi_{i})^{2}-\frac{2}{L^{2}}\cosh\phi_{i}\right)\right]\,,\label{eq:T_phi}\\
T_{\mu\nu}^{(F)} & =\sum_{\Lambda=1}^{4}Y_{\Lambda}\left(F_{\mu\rho}^{\Lambda}F^{\Lambda}{}_{\nu}{}^{\rho}-\frac{1}{4}g_{\mu\nu}F_{\rho\sigma}^{\Lambda}F^{\Lambda\rho\sigma}\right)\,.\label{eq:T_F} \\
Y_{i} & =e^{\phi_{1}+\phi_{2}+\phi_{3}-2\phi_{i}}\,,\qquad Y_{4}=e^{-\phi_{1}-\phi_{2}-\phi_{3}}\,,\qquad i=1,2,3\,.\label{eq:Y_def}
\end{align}
A solution of the above equations is a solution of the STU model whenever the following constraints, arising from setting the axion fields to zero in the full STU model, are satisfied
\begin{align}
e^{2\phi_{1}}F^{2}\wedge F^{3}-F^{1}\wedge F^{4} & =0\,,\nonumber \\
e^{2\phi_{2}}F^{1}\wedge F^{3}-F^{2}\wedge F^{4} & =0\,,\label{constraints_from_axiones}\\
e^{2\phi_{3}}F^{1}\wedge F^{2}-F^{3}\wedge F^{4} & =0\,.\nonumber 
\end{align}
These constraints are automatically fulfilled for purely electric or purely magnetic configurations. In the dilatonic sector, some simplifications occur in the supersymmetry transformations: the K\"ahler connection vanishes $\mathcal{Q}_{\mu}=0$,
the first four components of the $U(1)$-section are real $L^{\Lambda}\in\bR$,
and the superpotential is real $\mathcal{W}\in\bR$. Hence the Killing spinor equations, coming from the supersymmetry variations of the gravitini and gaugini, reduce to
\begin{align}
\delta\Psi_{\mu}^{A}\dd x^{\mu} & \equiv\mathfrak{D}^{A}(\epsilon)\label{KS equation chiral}
\\
 & \equiv\dd\epsilon^{A}+\frac{1}{4}\omega_{ab}\gamma^{ab}\epsilon^{A}+\frac{1}{2}A^{M}\theta_{M}\varepsilon^{AB}\delta_{BC}\epsilon^{C}+\frac{1}{4}L^{T}\mathcal{I}F_{ab}\gamma^{ab}\gamma\varepsilon^{AB}\epsilon_{B}+\frac{1}{2}\mathcal{W}\gamma\delta^{AB}\epsilon_{B}=0\,, \notag 
 \\
\delta\lambda^{iA} & =-\gamma^{\mu}\partial_{\mu}z^{i}\epsilon^{A}+\frac{1}{2}g^{i\overline{\jj}}\overline{f}_{\overline{\jj}}^{\Lambda}\mathcal{I}_{\Lambda\Sigma}F_{ab}^{\Sigma}\gamma^{ab}\varepsilon^{AB}\epsilon_{B}+g^{i\overline{\jj}}\overline{\mathcal{U}}_{\overline{\jj}}^{M}\theta_{M}\delta^{AB}\epsilon_{B}=0\,,
\end{align}
where we have defined the 1-form Clifford algebra valued $\gamma=\gamma_{a}e^{a}$. In this sector, the quantities $z^{i},\overline{f}_{\overline{\jj}}^{\Lambda}$
and $\overline{U}_{\overline{\jj}}{}^{M}\theta_{M}$ are purely imaginary. The quantity $\overline{\cal U}_{\overline{\jj}}{}^{M}$ is the U(1) covariant derivative of the U(1) section $\overline{V}^{M}$ and $\overline{f}_{\overline{\jj}}{}^{\Lambda}$ are the first four symplectic components of $\overline{\cal U}_{\overline{\jj}}{}^{M}$. The local supersymmetry parameters are chiral spinors satisfying $\gamma_{5}\epsilon_{A}=\epsilon_{A}$ and $\gamma_{5}\epsilon^{A}=-\epsilon^{A}$, and are related to each other through complex conjugation $(\epsilon^{A})^*=\epsilon_A$. $A^{M}$ are the symplectic gauge fields, and the embedding tensor defining the Fayet–Iliopoulos (FI) term, is given by
    \begin{align}
    \theta_{M} & =\frac{1}{\sqrt{2}L}(1,1,1,1,0,0,0,0)\,.
    \end{align}
Since we are considering real $\gamma$-matrices, all the coefficients in the above equation are real. We can write an equation for the Majorana spinor $\psi^{A}=\epsilon^{A}+\epsilon_{A}$
by combining (\ref{KS equation chiral}) with its complex conjugate. For simplicity, it is useful to define a complex spinor whose real and imaginary parts are the $SU(2)$ components of the Majorana spinor $\zeta=\psi^{1}+\ri\psi^{2}$. The Killing spinor equation for this complex Killing spinor and the gaugino equations, which are obtained by the same token, are given by
    \begin{align}
    \left(\dd+\frac{1}{4}\omega_{ab}\gamma^{ab}-\frac{\ri}{2}A^{M}\theta_{M}-\frac{\ri}{4}L^{T}\mathcal{I}F_{ab}\gamma^{ab}\gamma+\frac{1}{2}\mathcal{W}\gamma\right)\zeta & =0\, ,\label{eq zeta gravitino}
    \\
    \left(-\gamma^{\mu}\partial_{\mu}z^{i}-\frac{\ri}{2}g^{i\overline{\jj}}\overline{f}_{\overline{\jj}}^{T}\mathcal{I}F_{ab}\gamma^{ab}+g^{i\overline{\jj}}\overline{U}_{\overline{\jj}}{}^{M}\theta_{M}\right)\zeta & =0\,.\label{gaugino equations zeta}
    \end{align}
\section{4-Charge Soliton Configuration}\label{sec:soliton}
In this Section, we present a new solution of the STU model that describes a soliton. The four-charge, electrically charged black hole in the dilatonic sector of the STU model with a spherical horizon was originally constructed in \cite{Duff:1999gh}. Its extension to planar and hyperbolic horizon topologies was later obtained in \cite{Cvetic:1999xp}. Here we make a double Wick rotation of the planar solutions, which naively do not have a BPS limit. We find that the BPS limit of these configurations can be analyzed conveniently by introducing a parameter $q$ through an appropriate change of coordinates. We give the local form of the metric and the fields in coordinates $(t,r,\varphi,y)$. The metric is given by
    \begin{align}\label{solution}
    \dd s^{2} & =\frac{f(r)}{\sqrt{H(r)}}\dd\varphi^{2}+\frac{\sqrt{H(r)}}{f(r)}\dd r^{2}+r^{2}\sqrt{H(r)}(-\dd t^{2}+\dd y^{2})\,,\\
    f(r) & =\frac{r^{2}}{L^{2}}H(r)-\frac{m}{r}-\frac{q}{r^{2}}\,,\qquad H(r)=H_{1}H_{2}H_{3}H_{4}\,,\qquad H_{\Lambda}=1+\frac{q_{\Lambda}}{r}\,,\nonumber 
    \end{align}
and the dilatons and gauge fields are given by
    \begin{align}
    \phi_{1} & =\frac{1}{2}\log\left(\frac{H_{2}H_{3}}{H_{1}H_{4}}\right)\,,\qquad\phi_{2}=\frac{1}{2}\log\left(\frac{H_{1}H_{3}}{H_{2}H_{4}}\right)\,,\qquad\phi_{3}=\frac{1}{2}\log\left(\frac{H_{1}H_{2}}{H_{3}H_{4}}\right)\,,\nonumber \\
    A^{\Lambda} & =\left(\frac{Q_{\Lambda}}{\sqrt{2}rH_{\Lambda}}-\mu^{\Lambda}\right)\dd\varphi\,,\qquad Q_{\Lambda}^{2}=-q_{\Lambda}m+q\,. \label{matter fields sol}
    \end{align}
The configuration is controlled by the four parameters of the harmonic functions $q_\Lambda$, which are related to the four magnetic charges $Q_\Lambda$ through the last equation in \eqref{matter fields sol}, the parameter $m$ related to the energy of the configuration, and $q$ that was introduced through a shift in the radial coordinate and redefinition of the constants. The latter is relevant in the limit $m=0$, where the change of coordinates and the field redefinitions become singular, and therefore one cannot remove it once the limit is taken.  The range of the coordinates is $t,y\in \bR$, with $t$ being the timelike coordinate, $r\geq r_0$, where $r_0$ is a solution of the equation $f(r_0)=0$. When the above equation has a solution $r_0$ such that $r_0>-q_\Lambda$, then the configuration is free of curvature singularities, in particular is regular at $r=r_0$ if the coordinate $\varphi$ is periodic with a suitable period $\Delta$. 

Note that $\varphi$ is the coordinate of a contractible cycle that shrinks to zero when $r\to r_0$. This implies that the regularity of the 1-form gauge potentials requires the following relation between the parameters:
\begin{equation}
\mu^\Lambda = \frac{Q_\Lambda}{\sqrt{2} r_0 H_\Lambda(r_0)}\, .
\end{equation}

We shall use the following definition of the Wilson line at infinity
    \begin{align}
        \psi_\Lambda = \lim_{r\to \infty} \frac{1}{2 \pi} \oint A^\Lambda \implies \mu_\Lambda = -\frac{2 \pi \psi_\Lambda}{\Delta}
        \label{def psi Lambda}
    \end{align}
\subsection{Phase Space}
In this Section we will proceed by characterizing the phase space of the soliton solutions. This can be achieved by expressing the observables in terms of the boundary conditions, namely the boundary values of the gauge fields, $\mu_{\Lambda}$, and the periodicity of the coordinate $\varphi \in [0,\Delta]$. The latter is defined by the following regularity condition 
\begin{equation}\label{periodicity}
    \Delta = 4 \pi \nu \frac{\sqrt{H(r_0)}}{f'(r_{0})}
\end{equation}
where $\nu=\pm1$, in order to ensure $\Delta>0$.
In order to compute $f'(r_{0})$, we recall its definition for the four-charge soliton solution (\ref{solution})
\begin{equation}
    f(r) = \frac{r^{2}}{L^{2}}H(r) - \frac{m}{r} - \frac{q}{r^{2}}.
\end{equation}
 Defining
\begin{equation}
    h_{\Lambda} := r_{0}H_{\Lambda}(r_0) \, ,
\end{equation}
we can use the following  identities 
\begin{align}
    &H'_{\Lambda}(r) = \frac{1}{r}(1- H_{\Lambda}(r)) \, ,\\
    & f(r_{0}) = 0 \Rightarrow q=\frac{h_{1234}}{L^{2}} - r_{0}m \, ,\\
    & f'(r_{0}) = \frac{1}{r_{0}L^{2}}\left[h_{123} + h_{124}+h_{134}+h_{234} - m L^{2} \right] \, ,
\end{align}
 to rewrite (\ref{periodicity}) and solve it for $m$ 
\begin{equation}
    m = \frac{1}{L^{2}}\left[h_{123} + h_{124}+h_{134}+h_{234} \right] - \frac{4\pi \nu}{\Delta}\sqrt{h_{1234}} \, , \label{mass_solution}
\end{equation}
where we are adopting the following notation $h_{\Lambda\Sigma\dots\Theta} := h_{\Lambda}h_{\Sigma}\dots h_{\Theta}$.

Then, in order to relate the values at the horizon of the rescaled harmonic functions, $h_{\Lambda}$, to the boundary conditions $\mu_{\Lambda}$ we can consider the gauge fields of the solutions
\begin{equation}
    A^{\Lambda}  =\left(\frac{Q_{\Lambda}}{\sqrt{2}rH_{\Lambda}}-\mu^{\Lambda}\right)\dd\varphi
\end{equation}
together with the following relation, coming from the equations of motion
\begin{equation}
    Q_{\Lambda}^{2} = q - q_{\Lambda}m.
\end{equation}
This allows us to derive an important relation between  $\mu_{\Lambda}$, $h_{\Lambda}$ and the energy $m$ of the configuration
\begin{equation}\label{chemicalh}
    \mu_{\Lambda}^{2} = \frac{h_{1234}}{2 L^{2}h_{\Lambda}^{2}} - \frac{m}{2 h_{\Lambda}}\, .
\end{equation}
Finally, we can replace the expression  for $m = m (h_{\Lambda}, \Delta)$, derived in (\ref{mass_solution})
\begin{equation}
    \mu_{\Lambda}^{2} = \frac{1}{2L^{2}h_{\Lambda}}\left[\frac{h_{1234}}{h_{\Lambda}}- h_{123} - h_{124} -h_{134} - h_{234}\right] + \frac{2\pi\nu}{\Delta}\frac{\sqrt{h_{1234}}}{h_{\Lambda}}
\end{equation}
By explicitly computing them for $\Lambda = 1,2,3,4$, some partial simplification arises
\begin{align}\label{munonsuy}
    &\mu_{1}^{2} = -\frac{1}{2L^{2}}\left[h_{23} + h_{24} + h_{34}\right] + \frac{2\pi \nu}{\Delta}\frac{\sqrt{h_{1234}}}{h_{1}}\\
    &\mu_{2}^{2} = -\frac{1}{2L^{2}}\left[h_{13} + h_{14} + h_{34}\right] + \frac{2\pi \nu}{\Delta}\frac{\sqrt{h_{1234}}}{h_{2}}\\
    &\mu_{3}^{2} = -\frac{1}{2L^{2}}\left[h_{12} + h_{14} + h_{24}\right] + \frac{2\pi \nu}{\Delta}\frac{\sqrt{h_{1234}}}{h_{3}}\\
    &\mu_{4}^{2} = -\frac{1}{2L^{2}}\left[h_{12} + h_{13} + h_{23}\right] + \frac{2\pi \nu}{\Delta}\frac{\sqrt{h_{1234}}}{h_{4}}
\end{align}
In the above system of equations, we can linearly solve the first for $h_1$; then the system reduces to 

\begin{align}
   &\frac{h_{34}}{2 L^2} -\frac{8 \pi ^2 L^2 h_{34} \left(2 L^2 \mu_1^2+h_{34}\right)}{\Delta ^2 \left(2 L^2 \mu_1^2+h_{23}+h_{24}+h_{34}\right)^2}+\mu_2^2 = 0 \\
   & \frac{{h_{24}}}{2 L^2} -\frac{8 \pi ^2 L^2 h_{24} \left(2 L^2 \mu_1^2+h_{24}\right)}{\Delta ^2 \left(2 L^2 \mu_1^2 + h_{23}+h_{24}+h_{34}\right)^2}+ \mu_3^2 = 0\\
   &  \frac{h_{23}}{2 L^2}-\frac{8 \pi ^2 L^2 h_{23} \left(2 L^2 \mu_1^2+h_{23}\right)}{\Delta ^2 \left(2 L^2 \mu_1^2+h_{23}+h_{24}+h_{34}\right)^2}+\mu_4^2 = 0
\end{align}
One can decouple the algebraic system in terms of the variable
\begin{equation}
    Z=\frac{\Delta ^2 h_{34}}{4 \pi ^2 L^4}\,,
\end{equation}
a root of a certain polynomial $P(Z)$:
\begin{equation}\label{Zeq}
    P(Z)= 0
\end{equation}
The complete expression for this degree-$12$ polynomial $P(Z)$ is given in Appendix A. It is worth noting that, although obtaining analytic expressions for all the roots is computationally demanding, this polynomial in principle enables a complete characterization of the phase space for all non-supersymmetric solutions of the model.

\subsection{Supersymmetric configurations and scalars VEVs}
\begin{figure}
    \centering
    \includegraphics[width=1\linewidth]{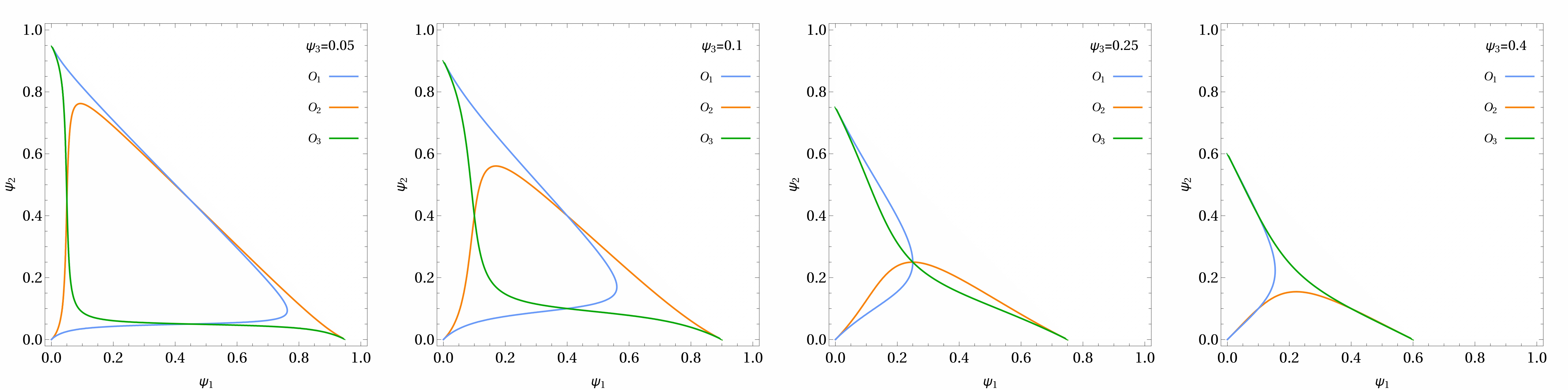}
    \caption{Locus in parameter space where the VEVs of the dual operators vanish, for different values of $\psi_3$, in units of $L$. The corresponding value of $\psi_4$ is determined from Eq. \eqref{psisum}.}
    \label{fig:vanishing_vevs}
\end{figure}

In this Section, we are interested in discussing the massless configurations $m=0$: this corresponds to BPS solutions, as we will show in the next Section. We notice that in this locus of the parameter space  the relations (\ref{chemicalh}) reduce to
\begin{equation}\label{chemicalhsusy}
    \mu^{2}_{\Lambda} = \frac{h_{1234}}{2 L^{2}h_{\Lambda}^{2}}
\end{equation}
and thus can be easily inverted and solved for the constants $h_{\Lambda}$
\begin{align}
   & h_{1}= L \sqrt{2\frac{\mu_{234}}{\mu_{1}}} = \sqrt{2}L |\mu_{1234}|^{\frac12} \frac{1}{|\mu_{1}|} \,,\\
   &h_{2}= L \sqrt{2\frac{\mu_{134}}{\mu_{2}}} = \sqrt{2}L |\mu_{1234}|^{\frac12} \frac{1}{|\mu_{2}|} \,, \\
   &h_{3}= L \sqrt{2\frac{\mu_{124}}{\mu_{3}}} = \sqrt{2}L |\mu_{1234}|^{\frac12} \frac{1}{|\mu_{3}|} \,, \\
   &h_{4}= L \sqrt{2\frac{\mu_{123}}{\mu_{4}}} = \sqrt{2}L |\mu_{1234}|^{\frac12} \frac{1}{|\mu_{4}|}\, ,
\end{align}
where again the notation used is  $\mu_{\Lambda\Sigma\dots\Theta} := \mu_{\Lambda}\mu_{\Sigma}\dots \mu_{\Theta}$. Interestingly, we can explicitly make use of the supersymmetric condition $m=0$ in \eqref{mass_solution}, in order to get an important relation
\begin{equation}
    |\mu_{1}|  + |\mu_{2}| + |\mu_{3}| + |\mu_{4}| = \frac{2\sqrt{2}\pi \nu L }{\Delta} 
\end{equation}
which can also be recast as 
\begin{equation}\label{psisum}
     |\psi_{1}|  + |\psi_{2}| + |\psi_{3}| + |\psi_{4}| = \sqrt{2} L 
\end{equation}
where we used the definition $\mu_{\Lambda} = -\frac{2\pi\psi_{\Lambda}}{\Delta}$ and set $\nu =1$, as it is the only consistent value.
At this stage, we have all the tools we need to compute the scalar VEVs. One can consider the scalars appearing in (\ref{matter fields sol}), together with the change of coordinates 
\begin{equation}
    \rho^{2} =r^{2}\sqrt{H(r)}
\end{equation}
which defines the radial coordinate in the $\mathrm{AdS}_{4}$ Poincaré patch. Then we can look at the asymptotic expansion $\rho \rightarrow \infty$
\begin{align}
    &\phi_{1}= \frac{-q_{1} +q_{2} + q_{3}-q_{4}}{2\rho} + O(\rho^{-2}) \, , \\
    &\phi_{2}= \frac{q_{1} -q_{2} + q_{3}-q_{4}}{2\rho} + O(\rho^{-2}) \, , \\
    &\phi_{3}= \frac{q_{1} +q_{2} - q_{3}-q_{4}}{2\rho} + O(\rho^{-2}) \, .
\end{align}
These relations define the VEVs of the dual operators as the coefficient of the leading order at the boundary (up to some relevant normalization).
\begin{align}
    &\mathcal{O}_{1}=\frac{-q_{1} +q_{2} + q_{3}-q_{4}}{2} \, , \\
    &\mathcal{O}_{2}=\frac{q_{1} -q_{2} + q_{3}-q_{4}}{2} \, , \\
    &\mathcal{O}_{3}=\frac{q_{1} +q_{2} - q_{3}-q_{4}}{2} \, .
\end{align}
Their can be expressed in terms of the $h_{\Lambda} = r_{0} + q_{\Lambda}$
   \begin{align}
    &\mathcal{O}_{1}=\frac{-h_{1} +h_{2} + h_{3}-h_{4}}{2}\\
    &\mathcal{O}_{2}=\frac{h_{1} -h_{2} + h_{3}-h_{4}}{2}\\
    &\mathcal{O}_{3}=\frac{h_{1} +h_{2} - h_{3}-h_{4}}{2}
\end{align}
and finally in terms of $\mu_{\Lambda}$ by means of (\ref{chemicalhsusy})
 \begin{align}
    &\mathcal{O}_{1}=\frac{L}{\sqrt{2}}|\mu_{1234}|^{\frac{1}{2}}\left(-\frac{1}{|\mu_{1}|}+\frac{1}{|\mu_{2}|}+\frac{1}{|\mu_{3}|}-\frac{1}{|\mu_{4}|}\right)\\
   &\mathcal{O}_{2}=\frac{L}{\sqrt{2}}|\mu_{1234}|^{\frac{1}{2}}\left(\frac{1}{|\mu_{1}|}-\frac{1}{|\mu_{2}|}+\frac{1}{|\mu_{3}|}-\frac{1}{|\mu_{4}|}\right)\\
    &\mathcal{O}_{3}=\frac{L}{\sqrt{2}}|\mu_{1234}|^{\frac{1}{2}}\left(\frac{1}{|\mu_{1}|}+\frac{1}{|\mu_{2}|}-\frac{1}{|\mu_{3}|}-\frac{1}{|\mu_{4}|}\right).
\end{align}
Equivalently,  a similar expression holds as functions of $\psi_{\Lambda}$
\begin{align}
    &\mathcal{O}_{1}=\frac{\sqrt{2}L}{\Delta}|\psi_{1234}|^{\frac{1}{2}}\left(-\frac{1}{|\psi_{1}|}+\frac{1}{|\psi_{2}|}+\frac{1}{|\psi_{3}|}-\frac{1}{|\psi_{4}|}\right)\\
   &\mathcal{O}_{2}=\frac{\sqrt{2}L}{\Delta}|\psi_{1234}|^{\frac{1}{2}}\left(\frac{1}{|\psi_{1}|}-\frac{1}{|\psi_{2}|}+\frac{1}{|\psi_{3}|}-\frac{1}{|\psi_{4}|}\right)\\
    &\mathcal{O}_{3}=\frac{\sqrt{2}L}{\Delta}|\psi_{1234}|^{\frac{1}{2}}\left(\frac{1}{|\psi_{1}|}+\frac{1}{|\psi_{2}|}-\frac{1}{|\psi_{3}|}-\frac{1}{|\psi_{4}|}\right).
\end{align}
Using (\ref{psisum}) one can actually remove the dependence on one of the $\psi_{\Lambda}$. The region in which each of the dual operators vanishes and the constraint \eqref{psisum} is satisfied corresponds to a codimension-2 surface in the 4-dimensional space parameterized by $(\psi_1, \psi_2, \psi_3, \psi_4)$. Different sections of these surfaces are shown in Figure \ref{fig:vanishing_vevs}.

\section{Supersymmetry analysis} \label{sec:susy}

To prove that the $m=0$ configuration preserves some amount of supersymmetry, we consider the following vielbein basis
    \begin{equation}
    e^0 = r H^{1/4} \dd t\, , \quad e^1=\frac{H^{1/4}}{f^{1/2}} \dd r \, , \quad e^2 = \frac{f^{1/2}}{H^{1/4}}\dd \varphi \, , \quad e^3 = r H^{1/4} \dd y \, .
    \end{equation}
The determinant of the matrix in \eqref{gaugino equations zeta}, coming from the spin 1/2, vanishes iff $m=0$. 

We shall show that this is a necessary condition to have a non-trivial Killing spinor. Indeed, the complex Killing spinor that solves \eqref{eq zeta gravitino} and \eqref{gaugino equations zeta} is 
\begin{align}
\zeta(r,\varphi) & =\frac12e^{-\ri\pi\varphi/ \Delta}\frac{f(r)^{1/4}}{H(r)^{1/8}}
\left[ \left(1 + \frac{\ri}{r}\sqrt{\frac{q} {f(r)}}\right)^{1/2} + \gamma^{1} \left(1 - \frac{\ri}{r}\sqrt{\frac{q} {f(r)}}\right)^{1/2} \right]
(1-\gamma^{2})\zeta_{0}\,, \label{complex KS} \, , \\
\Delta & = \frac{2\sqrt{2} \pi L}{
    \sum_\Lambda \mu^\Lambda} \, .
\end{align}
It is clearly anti-periodic in the coordinate $\varphi$, as should be because the coordinate $\varphi$ is contractible, and is well defined everywhere. The configuration is regular for $q>0$, which implies that $f(r)>0,H(r)>0$. The Majorana Killing spinor can be written in a simple way in terms of the real function
    \begin{align}
        X_\pm = \left( \pm 1 + \frac{r}{L}\sqrt{\frac{H(r)}{f(r)}} \right)^{1/2}
    \end{align}
which are related to complex functions as $\Re(Z_+)=X_{+}$, $\Im(Z_+)=X_{-}$. Majorana spinors transforms as a doublet of SO(2) and we denote them by $(\psi^A)=(\psi^1,\psi^2)^T=\underline{\psi}$. The Majorana Killing spinor associated with the complex Killing spinor \eqref{complex KS} is given by
    \begin{align}
        \underline{\psi} = \frac{f(r)^{1/4}}{H(r)^{1/8}} e^{\frac{\pi \varphi}{\Delta} \ri \sigma_2}[(1+\gamma^1) X_{+} - \ri \sigma_2 (1-\gamma^1) X_{-} ](1-\gamma^2)\underline{\psi_0} \, , 
    \end{align}
where $\sigma_2$ is the Pauli matrix that acts on the SO(2) indices and $\underline{\psi_0}$ denotes an arbitrary constant Majorana spinor.

\section{Conclusions}\label{sec:conclusions}

In this work we have constructed and analyzed a new family of supersymmetric AdS-soliton configurations in the purely dilatonic STU truncation of $D=4$ gauged $\mathcal{N}=8$ supergravity arising from the $S^7$ compactification of M-theory. The solutions are supported by four independent vector fields and provide everywhere regular horizonless geometries with a smoothly shrinking spatial circle and anti-periodic spin-structure along the contractible cycle.

A central result of the paper is the complete characterization of the supersymmetric moduli space. Imposing regularity at the infrared endpoint fixes the would-be integration constants in terms of the asymptotic Wilson lines, leaving a compact three-dimensional space of regular BPS vacua described by the constraint, which has the same structure than in five dimensions \cite{Anabalon:2026yxk},
\begin{equation}
|\psi_1|+|\psi_2|+|\psi_3|+|\psi_4|=\sqrt{2}L \, .
\end{equation}
Within this space, we derived explicit expressions for the vacuum expectation values of the three dimension-one scalar operators dual to the dilatons and showed that they provide a natural set of order parameters for the different supersymmetric branches.

The geometry of the moduli space exhibits a rich phase structure. The generic interior corresponds to regular supersymmetric vacua with all scalar condensates non-vanishing, while the codimension-two surfaces on which one of the VEVs vanishes, separate distinct Coulomb-branch-like sectors of the dual theory. Their intersections define loci of enhanced symmetry, and the fully symmetric point with equal Wilson lines corresponds to the AdS-soliton vacuum with vanishing scalar condensates. These results provide a precise holographic realization of the idea that infrared regularity dynamically selects the allowed supersymmetric vacua and fixes the scalar expectation values in terms of the boundary data.

We also solved the Killing spinor equations explicitly and constructed globally well-defined anti-periodic spinors throughout the spacetime. The analysis shows that the massless branch $m=0$ is precisely the supersymmetric locus of the general four-charge family and that the corresponding BPS solutions are free of curvature singularities and closed timelike curves in the physical region of parameter space.

From the holographic perspective, the solutions describe strongly coupled three-dimensional gauge theories compactified on a spatial circle in a confining phase. The scalar VEVs encode the pattern of infrared condensates, while the Wilson lines act as tunable boundary holonomies controlling the position in the supersymmetric moduli space. The existence of critical submanifolds where condensates vanish suggests the presence of non-trivial supersymmetric phase transitions analogous to those recently observed in other holographic Coulomb-branch and black-hole systems \cite{Anabalon:2025sok}.

Several directions deserve further investigation. It would be interesting to uplift the complete moduli space to eleven dimensions and determine the corresponding M2-brane distributions, to compute holographic observables such as Wilson loops and entanglement entropy across the critical surfaces, and to study the response of the supersymmetric solitons to finite temperature and angular momentum. More generally, our results indicate that regular supersymmetric AdS solitons with non-trivial vector fields provide a natural framework for exploring the interplay between confinement, moduli stabilization, and phase transitions in strongly coupled quantum field theories.

\section*{Acknowledgements}
The work of AA is supported in part by the FONDECYT grants 1230853, 1242043, 1250133, 1262452 and 1262414 and by the FAPESP grant 2024/16864-9.  MO is supported by AEI-Spain PID2023-152148NB-I00 and by Maria de Maeztu excellence unit grant CEX2023-001318-M, by Xunta de Galicia (CIGUS Network of Research Centres and project ED431F-2023/19), and by the European Union FEDER.

\appendix
\section{Non-supersymmetric configuration}
We give here the explicit expression for the polynomial $P(Z)$,  appearing in  equation (\ref{Zeq})
    \begin{align}
P(Z) = \Bigg( \bigg( \bigg( \bigg(\bigg( \bigg( \bigg( \bigg( ( 
        &p_{11}D_2 + p_{10} )D_1 + p_9 \bigg)D_2 + p_8 \bigg)D_1 +  \\
        &\quad + p_7 \bigg)D_2 + p_6 \bigg)D_1 + p_5 \bigg)D_2 + p_4 \bigg)D_1 + p_3 \bigg)D_2 + p_2 \Bigg)D_{1} + p_1(Z)
\end{align}
where $D_1 , D_2$ and $p_1,\dots,p_{11}$ are polynomials in the $Z$ variables, defined as follows 
\begin{align}
    D_{1} = (Z + 2 \psi_{1}^{2}) \hspace{0.5cm}   D_{2} = (Z + 2 \psi_{2}^{2}) 
\end{align}
and
\begin{align}
     p_1 = & -Z^5 D_{2}^{6}\\ \nonumber
    p_2  = & -Z^5 D_{1}^{5} \\ \nonumber 
    p_3 = &  \frac{1}{4}Z^{4}D_{2}^{4}\bigg(4 \left(\psi_2^4 - 2 \psi_2^2 \left(\psi_3^2 + \psi_4^2 - 1\right) + \psi_3^4 + 2 \psi_3^2 \left(\psi_4^2 - 1\right) + \psi_4^4 - 2 \psi_4^2 + 9\right) + \\ \nonumber  & +Z^2 + 4 Z \left(\psi_2^2 - \psi_3^2 - \psi_4^2 - 5\right)\bigg)\\ \nonumber
    p_4 = & \frac{1}{4}Z^{4}D_{1}^{3} \bigg(4 \left(\psi_1^4 - 2 \psi_1^2 \left(\psi_3^2 + \psi_4^2 - 1\right) + \psi_3^4 + 2 \psi_3^2 \left(\psi_4^2 - 1\right) + \psi_4^4 - 2 \psi_4^2 + 9\right) +\\
    \nonumber & +Z^2 + 4 Z \left(\psi_1^2 - \psi_3^2 - \psi_4^2 - 5\right)\bigg) \\
    \nonumber 
    p_5 = & \frac{1}{4}Z^3 D_{2}^2 \Bigg( 5 Z^3 + 2 Z^2 \left(-25 + 9 \psi_2^2 - 9 \psi_3^2 - 9 \psi_4^2\right) + 4 Z \left(28 + 3 \psi_2^4 + 5 \psi_3^4 + 6 \psi_3^2 \psi_4^2 + 5 \psi_4^4 - 8 \psi_2^2 \left(\psi_3^2 + \psi_4^2\right)\right) \\
    \nonumber &- 8 \bigg(\psi_2^6 + \psi_3^6 - \psi_3^4 \left(1 + \psi_4^2\right) + \left(-2 + \psi_4^2\right)^2 \left(3 + \psi_4^2\right) - \psi_2^4 \left(-5 + \psi_3^2 + \psi_4^2\right) - \psi_3^2 \left(8 - 18 \psi_4^2 + \psi_4^4\right) \\ \nonumber&- \psi_2^2 \left(-8 + \psi_3^4 + 8 \psi_4^2 + \psi_4^4 - 2 \psi_3^2 \left(-4 + \psi_4^2\right)\right)\bigg)\Bigg)\\ \nonumber
    p_6 =& \frac{1}{4}Z^3 D_{1}^2 \Bigg( 5 Z^3 + 2 Z^2 \left(-25 + 9 \psi_1^2 - 9 \psi_3^2 - 9 \psi_4^2\right) + 4 Z \bigg(28 + 3 \psi_1^4 + 5 \psi_3^4 + 6 \psi_3^2 \psi_4^2 + 5 \psi_4^4 - 8 \psi_1^2 \left(\psi_3^2 + \psi_4^2\right)\bigg) \\ \nonumber &- 8 \bigg(\psi_1^6 + \psi_3^6 - \psi_3^4 \left(1 + \psi_4^2\right) + \left(-2 + \psi_4^2\right)^2 \left(3 + \psi_4^2\right) - \psi_1^4 \left(-5 + \psi_3^2 + \psi_4^2\right) - \psi_3^2 \left(8 - 18 \psi_4^2 + \psi_4^4\right)\\ \nonumber & - \psi_1^2 \left(-8 + \psi_3^4 + 8 \psi_4^2 + \psi_4^4 - 2 \psi_3^2 \left(-4 + \psi_4^2\right)\right)\bigg)\Bigg)\\ \nonumber
     p_{7} =& \frac{41}{16} Z^6 + 2 Z^5 \bigg(-9 + 4 \psi_2^2 - 4 \psi_3^2 - 4 \psi_4^2\bigg) + \frac{1}{2} Z^4 \bigg(76 + 3 \psi_2^4 + 17 \psi_3^4 + 16 \psi_4^2 + 17 \psi_4^4 - 8 \psi_2^2 \Big(2 + 3 \psi_3^2 \\ \nonumber &  + 3 \psi_4^2\Big) + 2 \psi_3^2 \left(8 + 7 \psi_4^2\right)\bigg) - 4 Z^3 \bigg(2 \psi_2^6 + \psi_3^6 - \psi_3^4 \left(2 + \psi_4^2\right) + \left(-2 + \psi_4^2\right)^2 \left(2 + \psi_4^2\right) \\ \nonumber& - 2 \psi_2^4 \left(-2 + \psi_3^2 + \psi_4^2\right) - \psi_3^2 \left(4 - 20 \psi_4^2 + \psi_4^4\right) - \psi_2^2 \left(-4 + \psi_3^4 + 8 \psi_4^2 + \psi_4^4 - 2 \psi_3^2 \left(-4 + \psi_4^2\right)\right)\bigg) \\ \nonumber & + Z^2 \bigg(16 \psi_2^6 + \psi_2^8 - 2 \psi_2^4 \left(-16 + \psi_3^4 + 8 \psi_4^2 + \psi_4^4 - 2 \psi_3^2 \left(-4 + \psi_4^2\right)\right) + 8 \psi_2^2 \Big(\psi_3^4 + \left(-2 + \psi_4^2\right)^2 \\ \nonumber &  - 2 \psi_3^2 \left(2 + \psi_4^2\right)\Big) + \left(\psi_3^4 + \left(-2 + \psi_4^2\right)^2 - 2 \psi_3^2 \left(2 + \psi_4^2\right)\right)^2\bigg)
     \\ \nonumber
     p_8 = & \frac{1}{16} Z^2 \Bigg(41 Z^3 + 2 Z^2 \left(16 + 23 \psi_1^2 - 64 \psi_3^2 - 64 \psi_4^2\right) - 4 Z \bigg(64 + 17 \psi_1^4 - 10 \psi_3^4 - 80 \psi_4^2 - 10 \psi_4^4  \\ \nonumber &+ 20 \psi_3^2 \left(-4 + \psi_4^2\right) - 16 \psi_1^2 \left(-3 + \psi_3^2 + \psi_4^2\right)\bigg) + 8 \bigg(16 \psi_1^4 + \psi_1^6 - 2 \psi_1^2 \Big(-16 + \psi_3^4 + 8 \psi_4^2 + \psi_4^4  \\ \nonumber&- 2 \psi_3^2 \left(-4 + \psi_4^2\right)\Big) + 8 \Big(\psi_3^4 + \left(-2 + \psi_4^2\right)^2 - 2 \psi_3^2 \left(2 + \psi_4^2\right)\Big)\bigg)\Bigg)\\
     \nonumber 
     p_9 = & \frac{1}{4} Z \Bigg(11 Z^3 - 2 Z^2 \left(19 + 8 \psi_2^2 - 6 \psi_3^2 - 6 \psi_4^2\right) + 4 Z \bigg(12 + 10 \psi_2^2 + \psi_2^4 - \psi_3^4 - 8 \psi_4^2 - \psi_4^4 + 2 \psi_3^2 \left(-4 + \psi_4^2\right)\bigg) 
     \\ \nonumber &- 8 \bigg(4 \psi_2^2 + \psi_2^4 + \psi_3^4 + \left(-2 + \psi_4^2\right)^2 - 2 \psi_3^2 \left(2 + \psi_4^2\right)\bigg)\Bigg)\\ \nonumber 
     p_{10} = & \frac{1}{4} Z \left(-9 Z^2 - 4 \left(4 + \psi_1^2\right) + 2 Z \left(11 + \psi_1^2\right)\right) \\ \nonumber
     p_{11} = & 1 - Z + \frac{3}{8} Z^2
\end{align}

Even if we are not able to find radical forms of the roots for the most generic case, this polynomial factorizes in simpler forms in some truncations of the model. For example, we can show what happens when we restrict to the Einstein-Maxwell sector. This truncation is recovered when setting $h_{\Lambda} = h$ and $\psi_{\Lambda} = \psi$, which amounts to considering the solution (\ref{solution}) when all the scalar fields vanish and all the gauge fields are taken to be equal, as expected. In this case, the polynomial $P(Z) = 0$ reduces to
\begin{align}\label{polEM}
   P(Z) = (-1 + Z)^2 (Z + 2 \psi^2)^6 \left( Z(-4 + 9Z) + 12Z\psi^2 + 4\psi^4 \right) \left( Z^2 + 4\psi^4 - 4Z(1 + \psi^2) \right) = 0.
\end{align}
In order to analyze the consistency of the various branches that appear, one can use the definition of the variable $Z$
\begin{equation}
    Z = \frac{\Delta^2 h_{34}}{4 \pi^2 L^4}
\end{equation}
to solve for $\Delta$ using the different roots of (\ref{polEM}) and substitute it into (\ref{munonsuy}). Specifically, the solution $Z = 1$ should be excluded since it leads to $\psi^2 < 0$, as well as the two roots coming from the last factor, which lead to imaginary values for $\psi$. Finally, the two roots coming from the factor $\left( Z(-4 + 9Z) + 12Z\psi^2 + 4\psi^4 \right)$ produce two non-trivial solutions, requiring $0 \leq \psi^2 < \frac{1}{6}$.
\hypersetup{linkcolor=blue}
\phantomsection 
\addtocontents{toc}{\protect\addvspace{4.5pt}}
\bibliographystyle{mybibstyle}
\bibliography{biblio.bib}

\end{document}